\documentclass{sig-alternate-05-2015}

\usepackage{adjustbox}
\usepackage{multirow}
\usepackage{subcaption}
\usepackage[flushleft]{threeparttable}
\usepackage{xspace}
\usepackage{url}
\usepackage{comment}
\usepackage{listings}
\usepackage{color}
\usepackage{booktabs,siunitx}

\PassOptionsToPackage{hyphens}{url}\usepackage{hyperref}

\usepackage[shortlabels]{enumitem}

\newcommand{\system}{{\tt TrustShadow}\xspace}
\newcommand{\eg}{e.g.,}
\newcommand{\ie}{i.e.,}
\renewcommand{\paragraph}[1]{\vspace{1em} \noindent\textbf{#1}}
\newcommand{\losseparagraph}[1]{\vspace{0.5em} \noindent\textbf{#1}}

\lstdefinelanguage{armAssembler}{
    morekeywords={
        mov, movs, msr, cpsid, ldr, mrs, str, pop, eor, mcr, isb, movt
    },
    otherkeywords={
    },
    morecomment=[l]{//},
    morecomment=[l]{@},
    morestring=[b]"
    morestring=[b]'
}

\definecolor{codegreen}{rgb}{0,0.6,0}
\definecolor{codegray}{rgb}{0.5,0.5,0.5}
\definecolor{codepurple}{rgb}{0.58,0,0.82}
\definecolor{backcolour}{rgb}{0.95,0.95,0.92}
 
\lstdefinestyle{mystyle}{
    backgroundcolor=\color{backcolour},   
    commentstyle=\color{codegreen},
    keywordstyle=\color{magenta},
    numberstyle=\tiny\color{codegray},
    stringstyle=\color{codepurple},
    basicstyle=\footnotesize\ttfamily,
    breaklines=true,                 
    captionpos=b,                    
    keepspaces=true,
    otherkeywords={mov,movs},       
    numbers=left,                    
    numbersep=5pt,                  
    showspaces=false,                
    showstringspaces=false,
    showtabs=false,                  
    tabsize=2
}
\begin{document}
\sloppy

\CopyrightYear{2017} \setcopyright{acmcopyright}
\conferenceinfo{MobiSys '17,}{June 19--23, 2017, Niagara Falls, NY, USA.}
\isbn{978-1-4503-4928-4/17/06}\acmPrice{\$15.00}
\doi{http://dx.doi.org/10.1145/3081333.3081349}

\clubpenalty=10000 
\widowpenalty = 10000







\title{TrustShadow: Secure Execution of Unmodified Applications with ARM TrustZone}

\numberofauthors{1}
\author{
\alignauthor 
        {Le Guan$^{\ast}$, Peng Liu$^{\ast}$, Xinyu Xing$^{\ast}$, Xinyang Ge$^{\#}$, \\ Shengzhi Zhang$^{\dagger}$,  Meng Yu$^{\ddagger}$, Trent Jaeger$^{\ast}$}\\
        \affaddr{$^{\ast}$Pennsylvania State University}  \\
        \affaddr{\{lug14, pliu, xxing\}@ist.psu.edu,
        \{tjaeger\}@cse.psu.edu} \\
        \affaddr{$^{\#}$Microsoft Research} \\
        \affaddr{\{xing\}@microsoft.com} \\
        \affaddr{$^{\dagger}$Florida Institute of Technology} \\
        \affaddr{\{zhangs\}@cs.fit.edu} \\
        \affaddr{$^{\ddagger}$University of Texas at San Antonio} \\
        \affaddr{\{meng.yu\}@utsa.edu}
}

\maketitle


\begin{abstract}


The rapid evolution of Internet-of-Things (IoT) technologies has led to an emerging need to make them smarter. 
A variety of applications now run simultaneously on an ARM-based processor.
For example, devices on the edge of the Internet are provided with higher horsepower to be entrusted with storing, processing and analyzing data collected from IoT devices.
This significantly improves efficiency and reduces the amount of data that needs to be transported to the cloud for data processing, analysis and storage.
However, commodity OSes are prone to compromise. Once they are exploited, attackers can access the data on these devices. Since the data stored and processed on the devices can be sensitive, left untackled, this is particularly disconcerting.

In this paper, we propose a new system, \system that shields legacy
applications from untrusted OSes. \system takes advantage of ARM TrustZone technology and
 partitions resources into the secure and normal worlds. In the secure world, \system constructs a trusted execution environment 
 for security-critical applications. This trusted environment is
maintained by a lightweight runtime system that coordinates the communication
between applications and the ordinary OS running in the normal world. The
runtime system does not provide system services itself. Rather, it forwards
requests for system services to the ordinary OS, and verifies the correctness
of the responses. To demonstrate the efficiency of this design, we prototyped
\system on a real chip board with ARM TrustZone support, and evaluated its
performance using both microbenchmarks and real-world applications. We showed
\system introduces only negligible overhead to real-world applications.

\end{abstract}

%


%
%

%
%
\printccsdesc




\section{Introduction}
\label{sec:intro}

The emerging Internet of Things (IoT) technologies have
enabled more and more isolated ``things'' to collect, process,
analyze, and exchange data. To become smart,
IoT devices tend to be built atop ARM-based multi-programming platforms,
in which multiple programs run simultaneously on commodity Operating Systems (OSes).
This allows them to install product-ready applications and 
saves effort and budget in application development.


For example,
edge computing~\cite{fogcomputing,foganalytics} is a decentralized computing infrastructure
that connects IoT devices to process their data on other,
more powerful devices that are on -- or close to -- the network edge. Examples
of such edge devices include industrial controllers, smart gateway and
routers, embedded servers, and even automotive in-vehicle infotainment etc. Compared to a central data center in the cloud, edge devices are
geographically closer to the IoT devices. Thus, they can respond to IoT
devices more quickly, making them more suitable to environments where operations
are time-critical or Internet connectivity is poor. 
In telemedicine scenarios, edge devices could run sophisticated analytics and 
turn around results in real time~\cite{DBLP:journals/corr/ConstantBADM17}.
As another example, a smart 3D-printer~\cite{3dprintingfirstv7} can directly
download an STL (STereoLithography) file from the ``MakerBot Digital Store''~\cite{3dprintingfirstv7}, convert it into \texttt{GCODE} by running a slicer program,
and feed the \texttt{GCODE} to the actuator.



Unfortunately, the security
provided by commodity OSes in the multi-programming platforms is often inadequate. Once an OS is compromised,
attackers gain complete access to the data on a system. Since these devices
may often deal with confidential data, possibly subject to laws and
regulations, this is particularly disconcerting.

To address this problem, a straightforward reaction is to safeguard
applications against the OSes potentially vulnerable to exploitable bugs or
misconfiguration. Prior efforts on this explore executing applications
that handle sensitive data in separate virtual machines
(e.g.,~\cite{chen2008overshadow,hofmann2013inktag,proxos}), taking advantage
of hardware features (e.g.,~\cite{BaumannPeinadoHunt2014,flicker,trustvisor})
or retrofitting commodity OSes (e.g.~\cite{CriswellDautenhahnAdve2014}).
Unfortunately, these solutions are not applicable to the aforementioned IoT multi-programming platforms.


First, these devices do not have the hardware features typically available on
PCs. To be energy efficient, these devices generally incorporate ARM Cortex-A
processors, making the techniques that rely on unique hardware completely
futile (e.g., Haven~\cite{BaumannPeinadoHunt2014} based on Intel SGX). Second,
these devices do not have abundant computational resources in comparison with
PCs or a data center in the cloud. Thus, it is not realistic to adopt to these devices those
techniques specifically designed for PCs or data centers. Last but not least,
some techniques previously proposed require radical changes to applications
and OSes, which poses a substantial barrier to their adoption. This is
especially true in the scenario where device manufacturers would like to
retain compatibility with existing applications.

In this paper, we address the aforementioned issues by developing \system.
\system is a system that shields legacy applications from a compromised OS. By
taking advantage of ARM TrustZone technology~\cite{tz}, our system constructs
a trusted execution environment for security-critical applications. Different from some existing techniques, \system does not radically
change existing OSes. Rather, it utilizes a lightweight runtime system to
coordinate communications between applications and untrusted OSes. As such,
\system requires no changes to existing applications either.


More specifically, we develop \system with a runtime system running in the
TrustZone of an ARM processor. The runtime manages the page tables for
applications locally in an isolated secure environment, and ensures their virtual
memory cannot be accessed by an untrusted OS running outside the environment.
To accommodate the execution of applications in a lightweight manner, the
runtime does not incorporate complicated system services. Rather, it forwards
application requests for system services to
the untrusted OS, similar to Proxos~\cite{proxos}. To guarantee
security, the runtime verifies
return values from system services to defeat Iago attacks~\cite{iago}, and interposes
context switches between the applications and the untrusted OS. Considering
an application might interact with file I/O, the runtime system also encrypts the
data before revealing it to the untrusted OS for storing.

With the design above, \system protects legacy applications from the untrusted
OSes running them. As a result, developers no longer need to re-engineer
applications in order to run them on IoT devices. Since \system does not
implement system services itself, the complexity of Trusted Computing Base
(TCB) is reduced, making \system less vulnerable to exploits. To the best of
our knowledge, \system is the first solution on ARM-based IoT devices that allows an
unmodified application 
to run protected from attacks from untrusted OSes

In summary, this paper makes the following contributions.

\begin{itemize} 
	\item We propose a system -- \system -- for ARM-based multi-programming platforms. It can protect security-critical applications from untrusted OSes without the requirement of re-engineering the applications.

	\item We introduce a runtime system within \system. It accommodates the execution of Linux applications with a lightweight forwarding-and-verifying mechanism.

	\item We implemented \system on a real chip (SoC) board with the ARM TrustZone support with only about 5.3K lines of code (LOC) in the secure world, and 300 LOC in the normal world. Using microbenchmarks and real world software, we showed that \system imposes only negligible performance overhead.

\end{itemize}

The rest of the paper is organized as follows. Section~\ref{related} and~\ref{preliminary} present related work and the background of TrustZone, respectively. Section~\ref{sec:threatmodelandgoals} discusses our threat model. Section~\ref{sec:overview} describes the overview of \system. Section~\ref{sec:design} and~\ref{implementation} introduce our design and prototype implementation in detail. We present the evaluation of \system in Section~\ref{sec:eval}, followed by some discussion in Section~\ref{sec:discussion}. Finally, we conclude the paper in Section~\ref{sec:conclusion}.

\section{Related Work}
\label{related}

As is described in Section~\ref{sec:intro}, prior research primarily focuses on taking advantage of virtual machines, hardware features and radical code re-engineering to protect applications from compromised OSes. In this section, we summarize these works and describe why they are not suitable for IoT devices with more details.




\paragraph{Hypervisors and Virtual Machines.}
To protect an application from a compromised OS, one research effort focuses on utilizing hypervisor to construct trusted execution environment for applications. Systems following this design principle include {\tt Overshadow}~\cite{chen2008overshadow}, {\tt CHAOS}~\cite{ChenZhangChenEtAl2007}, {\tt SP$^3$}~\cite{YangShin2008}, {\tt Inktag}~\cite{hofmann2013inktag}, etc. They encrypt address space for an application under protection through a hypervisor, so that a compromised OS can only view the address space of the application in ciphertext. Using the hypervisor, they also verify the integrity of memory contents, and thus ensure a compromised OS cannot jeopardize the execution of the application. Similar to these techniques, another research effort focuses on escalating protection with virtual machines. For example, {\tt Terra}~\cite{terra} and {\tt Proxos}~\cite{proxos} allocate a dedicated VM for an application, making it resistant to a malicious OS. 

While these systems have been shown to be effective in shielding applications, they are an overkill for resource-constrained IoT devices, and sometimes
cannot be adopted by IoT devices. 
First, deploying a hypervisor-based system cannot provide the best (native) performance for the already performance-hungry IoT devices~\cite{nohyperviosr}.
Second, virtualization extension used in {\tt InkTag} etc., is a new hardware feature for the ARM platform, and is missing for many existing ARM devices\footnote{ARM released virtualization extension in the year 2010~\cite{armhyper}.}. 
Third, ARM has recently released the new IoT-oriented Cortex-M processor series which incorporate TrustZone extension, not virtualization extension~\cite{cortexm}.
This meets our speculation that virtualization is not suitable for resource-constrained IoT devices.



From the security perspective, hypervisor or virtual machine based solutions
relies on hypervisor, which is already struggling with its own security
problems due to increasing TCB size~\cite{xenbug,vmwarebug}. In this work,
\system harnesses TrustZone technology to mediate communication between OS and
applications, which eliminates complex, error-prone resource allocation in a
hypervisor.

\paragraph{Hardware Features.}
Research in the past also explores using various hardware features to protect applications from untrusted OSes. For example, {\tt Haven}~\cite{BaumannPeinadoHunt2014} takes advantage of Intel Software Guard eXtension (SGX)~\cite{sgx} to safeguard applications. More specifically, it harnesses SGX to instantiate a secure region of address space, and then protects execution of applications within that region from malicious privilege code. In addition to Intel SGX, Trusted Platform Module (TPM) is also used for shielding applications from a potentially malicious OS. For example, both {\tt Flicker}~\cite{flicker} and {\tt TrustVisor}~\cite{trustvisor} utilize TPM to isolate the execution of sensitive code. As is described in Section~\ref{sec:intro}, IoT devices generally incorporate ARM Cortex-A processors which do not have the aforementioned hardware features. As a result, previous techniques based on those cannot be applicable.

Trusted Language Runtime ({\tt TLR})~\cite{santos2014using}, {\tt
VeriUI}~\cite{veriui} and {\tt TrustOTP}~\cite{trustotp} utilize ARM TrustZone
technology for shielding applications. {\tt TLR} 
implement a
small runtime capable of interpreting .NET managed code inside
the secure world. By splitting mobile application into secure part and non-secure part,
the secure part of the app is never exposed to the untrusted OSes.
{\tt VeriUI} utilizes TrustZone to provide a trustworthy setting
for handling passwords. {\tt TrustOTP} harnesses TrustZone to protect the
confidentiality of the One-Time-Password against a malicious mobile OS. While
these works take advantage of TrustZone, they require modifications to
applications in order to be under protected. This poses a substantial barrier to their adoptions.


\paragraph{Code Instrumentation.}
{\tt Virtual Ghost}~\cite{CriswellDautenhahnAdve2014} is another research endeavor on protecting applications from a hostile OS. Different from those techniques discussed above, it uses compiler techniques and run-time checking to implement a mechanism similar to {\tt InkTag} within the OS kernel. Since the compiler instrumentation and run-time checking introduce more privilege code to kernel, not only does it increase TCB of a computer system but also imposes performance overhead, making it not suitable to energy-efficient, computation-lightweight IoT devices.

\section{TrustZone}
\label{preliminary}

In this section, we present the background of ARM TrustZone technology. To be more specific, we briefly describe its architecture, address space
controller and memory management unit (MMU).

\subsection{Architecture}

ARM TrustZone partitions all of the System-on-Chip (SoC) hardware and software resources in one of two worlds - the secure world for the security subsystem, and the normal world for everything else. With this partition, a single physical processor core can safely and efficiently execute code from both the normal world and the secure world in a time-sliced fashion. When the processor executes code in the normal world, it enters a non-secure state in which the processor can only access resources in the normal world. Otherwise, it is in a secure state in which the processor can access resources resided in both worlds.

To determine whether program execution is in the secure or normal world, ARM TrustZone extends a Non-Secure bit (NS-bit) on the AMBA Advanced eXtensble Interface (AXI) bus. With this NS-bit, the processor can check permissions on the access. To manage switches to and from the secure world, TrustZone provides \emph{monitor mode software} which ensures the state of the world that the processor is leaving is safely saved, and the state of the world the processor is switching to is correctly restored. The secure world entry to the monitor mode can be achieved by an explicit call via an {\tt smc} instruction.

\subsection{Address Space Controller}

TrustZone Address Space Controller (TZASC) is an Advanced Microcontroller Bus Architecture (AMBA) compliant SoC peripheral. It allows a TrustZone system to configure security access permissions for each address region. In TrustZone, the access permissions are managed by a group of registers, the access to which must be from the secure world. In addition, TZASC controls data transfer between an ARM processor and Dynamic Memory Controller (DMC). To permit data transfer, it examines whether NS-bit matches the security settings of the memory region. Given a memory region set to \emph{secure access only}, for example, an attempt to read returns all zeros and that to write has no change to the contents in that region.  




\subsection{Memory Management Unit}

An ARM processor also provides MMU to perform the translation of virtual memory addresses to physical addresses. Since TrustZone partitions memory space into secure and normal worlds, a processor with TrustZone enabled provides two separated virtual MMUs which allow each world to map virtual addresses to physical addresses independently. 

In the normal world, a process can only access physical memory in the non-secure state. In the secure world, it however can specify how to access physical memory by tuning NS-bit. For example, it could adjust the NS field in the first-level page table, and access the memory in either the secure or non-secure state. This flexibility augments a TrustZone system with an ability to efficiently share memory across the worlds.



\section{Threat Model}

\label{sec:threatmodelandgoals}
\label{sec:threatmodel}


\system shields a trustworthy application from a hostile OS. We consider a
completely compromised OS, which means the attacker can execute arbitrary
hostile code with system privilege to interfere with the memory and registers
of a process. For example, it may read/write any memory in victim process's
address space, through either load/store instructions or Direct Memory
Access~\cite{DMA2, DMA4}, causing memory disclosure and code injection
attacks. As another example, OS could modify interrupted process state
(\eg~the \texttt{PC} register) during exception handling and resume the
execution from an arbitrary instruction to change the program execution's
control flow.

In addition, the OS could change victim process's behavior by hijacking system
services (\eg~forging system call responses), leading to Iago
attacks~\cite{iago}. Recent study shows that an adversary can infer data by
observing a program's page fault patterns~\cite{controlled}. This kind of
controlled channel attack is also covered in this work.

Availability is out of scope in this paper.
In fact, a compromised OS could simply refuse to boot, or decline
the time slices of an HAP to launch Denial-of-Service (DoS) attacks.
Side channel attacks such as timing and power analysis are out of scope in this paper.
We assume the runtime system running in the TrustZone is trusted. Throughout
our design of \system, we keep its functionality simple and its code base
minimal. This makes it easier to ensure its correctness through formal
verification~\cite{formalverification} or manual review.

\newlist{PR}{enumerate}{1}
\setlist[PR]{label=\textbf{S\arabic*.}}


\begin{figure}[t]
        \centering
        \includegraphics[width=0.9\columnwidth]{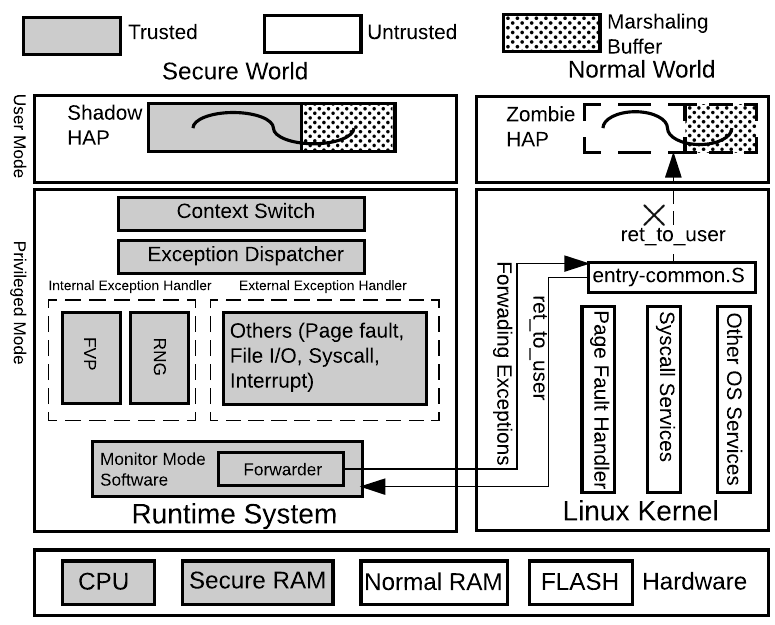}
        \caption{The architecture of \system.}
        \label{fig:overview}
\end{figure}

\section{Overview}
\label{sec:overview}

Figure~\ref{fig:overview} illustrates the architecture of \system, where the runtime system and Linux kernel run in the secure and normal world, respectively. Within the secure world, the runtime system shields the execution of a High-Assurance Process (HAP), and all the trusted modules shown in the figure cannot be accessed by the ordinary Linux running in the normal world. 


To be resistant to a hostile OS, an application needs to be initialized through a customized system call, which creates a ``zombie'' HAP and its ``shadow'' counterpart. In our design, the zombie HAP represents the application running in the normal world. However, it \emph{never} gets scheduled to run. Rather, \system runs its ``shadow'' counterpart residing in the secure world. To support the execution of the shadow HAP, \system introduces a lightweight runtime system to the secure world.


The runtime system does not provide system services for shadow HAPs. Instead, it intercepts exceptions and forwards them to the Linux OS running in the normal world. In this way, the runtime system can maintain a trusted execution environment for HAPs without introducing a large amount of code to the secure world. To enable cross-world communications, \system introduces data structure {\tt task\_shared} to share data between the runtime system and the OS. In addition, \system sets aside data structure {\tt task\_private} to store sensitive metadata for shadow HAPs. In our design, this data structure can only be accessed by the runtime system.

To accommodate the execution of HAPs and coordinate communications across two worlds, the runtime system is designed with various modules (see Figure~\ref{fig:overview}). Serving as the gateway for all exceptions and their returns, the context switch module maintains the CPU hardware context for each HAP, and restores/clears general-purpose registers accordingly. It allows the runtime system to coordinate the execution of an HAP and avoid leaking sensitive data to the OS running in the normal world.

As is shown in Figure~\ref{fig:overview}, the runtime system also implements an internal exception handler module -- indicated by FVP and RNG. They are designed to handle floating point computation and random number requests locally, for the reasons that cryptographic operation must rely upon trustworthy random number generation, and floating point computation necessarily exposes floating registers if \system relies upon the Linux OS for handling this exception. In Section~\ref{sec:localservice}, we describe this internal exception handler in detail.

In our design, the runtime system handles exceptions using three modules,
including exception dispatcher, external exception handler, and forwarder. The
exception dispatcher is responsible for dispatching exceptions to
corresponding handlers. Except for floating point exception and random number
requests, this module dispatches all the exceptions to the external exception
handler which further redirects the exceptions to the forwarder module. To
accommodate exception forwarding in a transparent manner, the forwarder module
\emph{emulates} an exception context for the normal world, pretending that
exception is trigger by the zombie HAP. After receiving exceptions, the Linux
OS handles them and returns results through {\tt task\_shared}. The external exception
handler verifies the return results before reflecting them to the execution
environment of the corresponding HAP. In Section~\ref{sec:forward},~\ref{sec:pagetablesyn} and~\ref{sec:intervention}, we describe how the three modules coordinate and perform external exception handling. 




Since the normal OS does not have the privilege to access a shadow HAP, \system also introduces a world-shared buffer, indicated as the marshaling buffer in Figure~\ref{fig:overview}. Through this buffer, not only does \system share the parameters of system calls with the ordinary OS but also retrieves the returns of the system calls. To retrieve the return value of a system call, \system copies the data in the buffer to the memory region corresponding to the system call, provided that the verifier module marks it valid.

\section{Runtime System}
\label{sec:design}


In this section, we detail the runtime system illustrated in Figure~\ref{fig:overview}. We begin with memory management for security. Then, 
we discuss how the aforementioned modules coordinate HAP execution. More specific, we describe how they forward exceptions, handle page faults and intervene system calls. 
Last, we present internal exception handling.

\begin{figure}
        \centering
        \includegraphics[width=\columnwidth]{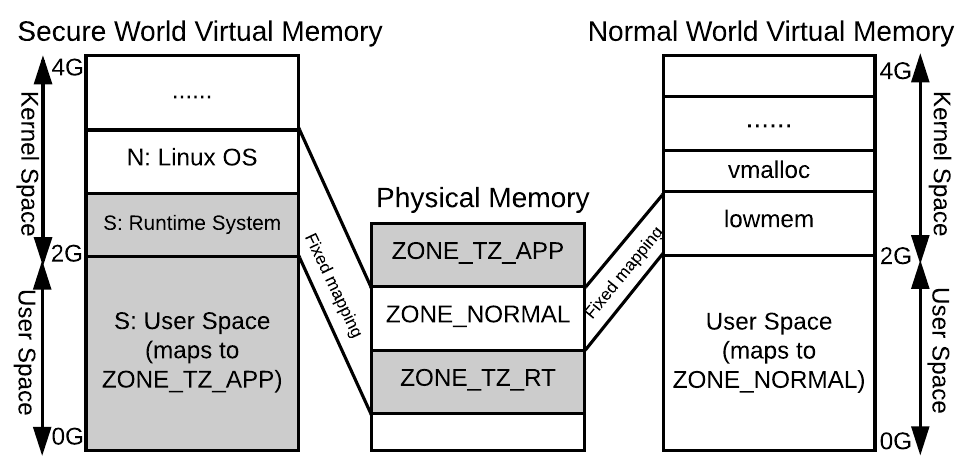}
        \caption{Physical memory partition vs. virtual memory layout.}
        \label{fig:asmap}
\end{figure}

\subsection{Memory Management}

Here, we describe how we partition physical memory regions, and specify the
design of virtual memory system.




\paragraph{Physical Memory Partition.}
\label{memoryisolation}
Using TZASC, \system creates three distinct physical memory regions. They are non-secure region {\tt ZONE\_TZ\_APP} as well as secure regions {\tt ZONE\_TZ\_RT} and {\tt ZONE\_TZ\_APP}. The non-secure region can be accessed by both the normal and secure worlds, whereas the secure regions have to be accessed through the secure world. In our design, we designate secure region {\tt ZONE\_TZ\_APP} for shadow HAPs, {\tt ZONE\_TZ\_RT} for the runtime system and non-secure region {\tt ZONE\_NORMAL} for holding the Linux OS and other ordinary processes. We illustrate these three regions in Figure~\ref{fig:asmap}.
With the partition above, the runtime, HAPs and Linux OS are all physically isolated, which provides the essential support for safeguarding the HAPs. 


\paragraph{Virtual Memory Layout.}
\label{addressspace}
\system supports executing legacy Linux code in the secure world. As a result,
we design the virtual address of the secure world to follow the same
user/kernel memory split as that in the Linux OS. With this design, legacy
code can be offloaded to execute in the secure world without any code
relocation. In our current design, both Linux OS and the runtime system
maintain a 2G/2G virtual address split, as shown in Figure~\ref{fig:asmap}.

In the kernel space of secure world, in addition to mapping itself to {\tt
ZONE\_TZ\_RT}, the runtime system maps the physical memory holding the Linux
OS ({\tt ZONE\_NORMAL}) in the virtual address space.  With this mapping, the
runtime system can efficiently locate shared data from the OS (such as {\tt
task\_shared}) by adding a corresponding offset.




\subsection{Forwarding Exceptions}
\label{sec:forward}

In general, a program is not self-contained. During execution, it might be trapped into the OS (e.g., calling a system service, encountering a page fault or interrupt). In the ARM architecture, system calls are requested by issuing the \texttt{svc}
instruction which traps the processor into privileged \texttt{SVC} mode to accomplish the system services. Likewise, other exceptions during execution would trap the processor into the corresponding privileged modes.

As is described in Section~\ref{sec:overview}, except for float point computation and random number generation, the runtime system intercepts exceptions and redirects them to the Linux running in the normal world. Here, we describe how \system performs exception forwarding. 


ARM processors utilize current program status register (\texttt{cpsr}) to hold
the current working mode of a processor (e.g., \texttt{USR} or \texttt{SVC}).
When an exception is taken, a processor enters the target mode by performing
the following operations. First, register \texttt{pc} points to the
corresponding offset in the exception vector table. Then, the processor stores
the value of previous \texttt{cpsr} to saved program status register
(\texttt{spsr}) before setting \texttt{cpsr} to indicate the current working
mode (\ie~the target mode). In the ARM architecture, \texttt{spsr} is a banked
register and thus each processor mode has its own copy. Based on the value of
\texttt{spsr}, an exception handler could get information about the
pre-exception processor mode.

Since the monitor mode software can access resources in both worlds, the
runtime system can \emph{re-produce} an exception as follows. Here, we take
forwarding a \texttt{SVC} exception as an example. (i) The runtime system sets
\texttt{spsr} in monitor mode to represent the target mode (\texttt{SVC}).
(ii) It sets the target mode's \texttt{spsr} to represent user mode
(\texttt{USR}). (iii) It issues \texttt{movs} instruction to jump to the
target exception handler (\texttt{0xFFFF0008}). Here, \texttt{movs} is an
exception return instruction. In addition to jumping to the target address, it
copies \texttt{spsr}  in the current mode (\texttt{SVC}, which is set in Step
i) to \texttt{cpsr} in the target mode. As a result, the OS kernel catches the
   exception at the correct address (\texttt{0xFFFF0008}) in the right mode
   (\texttt{SVC}), with \texttt{spsr} indicating that the exception comes from
   user mode (set in Step ii). We provide a code snippet to demonstrate this
   implementation in Appendix~\ref{appen:forwarding}. Forwarding other types
   of exception can be implemented in a similar way.

\subsection{Handling Page Fault}
\label{sec:pagetablesyn}
A page fault is one type of exception resulting from the failure of fetching an instruction or accessing data.
In the ARM architecture, a page fault is also called an abort exception, raised by MMU, indicating that the memory accessed
does not have a page table entry set properly. After such an exception is taken, an OS invokes its page fault handler which assigns an appropriate physical page and updates the page table entry accordingly. Typically, a page table
entry includes the virtual-to-physical address mapping and the access
permissions of the virtual memory.


In general, an OS maintains page tables for applications. However,
considering that an OS might be hostile and can tamper with the page tables
for applications, we isolate these page tables from the OS by placing them in
the secure world. The runtime system updates their entries by taking advantage
of the page fault handler provided by Linux OS.

To harness the page fault handler, we modify the existing on-demand page fault
handing mechanism in Linux. In particular, we hook the page fault handler so
that it can store the context of page fault handling in the aforementioned
shared memory, \texttt{task\_shared}\footnote{In our design, {\tt
task\_shared} carries the updated page table entry value (which contains the
address of the translated physical memory page), the influenced virtual
address, and additional contextual information. }. After retrieving the
updating information and before installing a page table entry, the runtime
system validated the returned information. In the following, we provide more
details on this procedure.




\begin{figure*}[t!]
    \centering
    \begin{subfigure}[t]{0.32\textwidth}
        \centering
        \includegraphics[width=1\columnwidth]{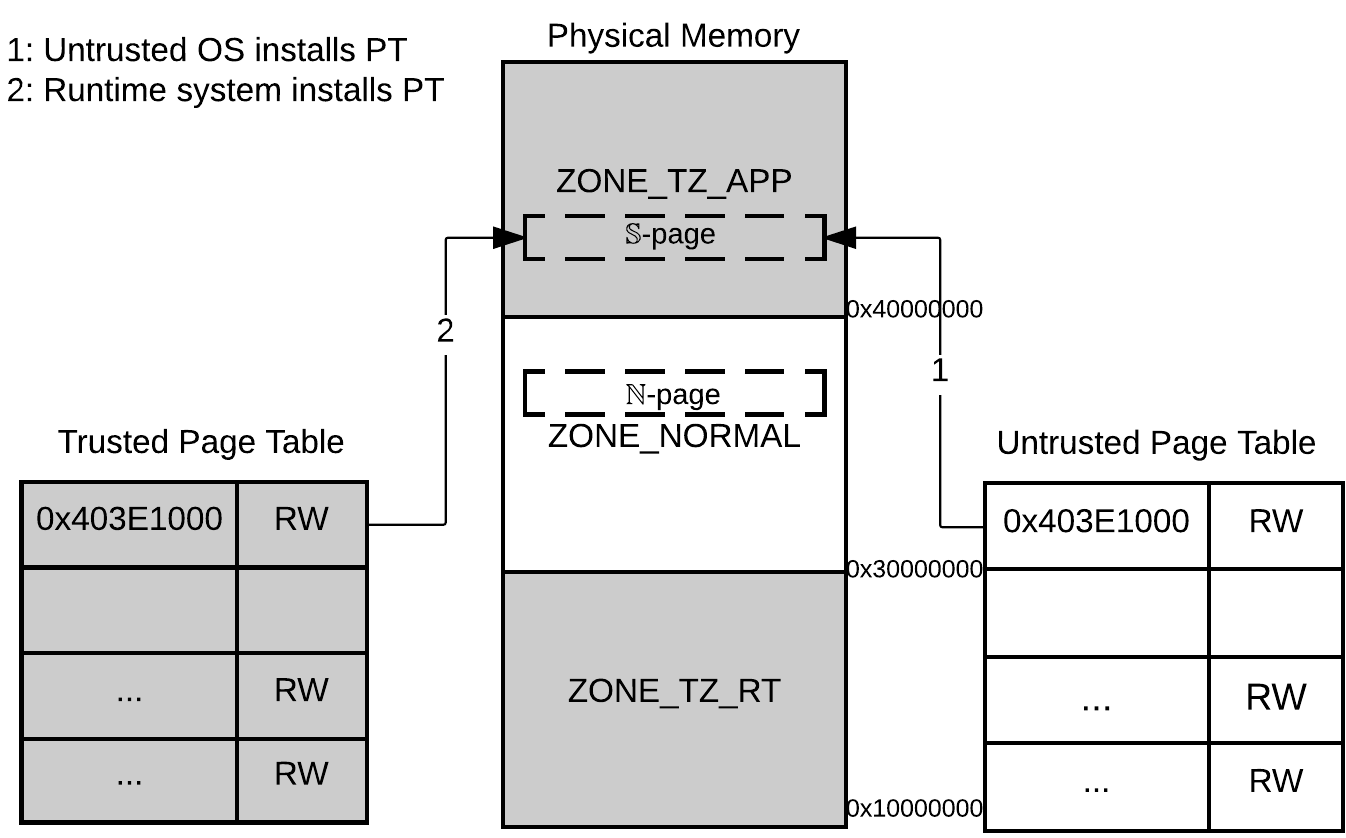}
        \caption{Basic Page Table Update.}
        \label{fig:syncpt1}
    \end{subfigure}%
    ~   
     \begin{subfigure}[t]{0.32\textwidth}
        \centering
        \includegraphics[width=1\columnwidth]{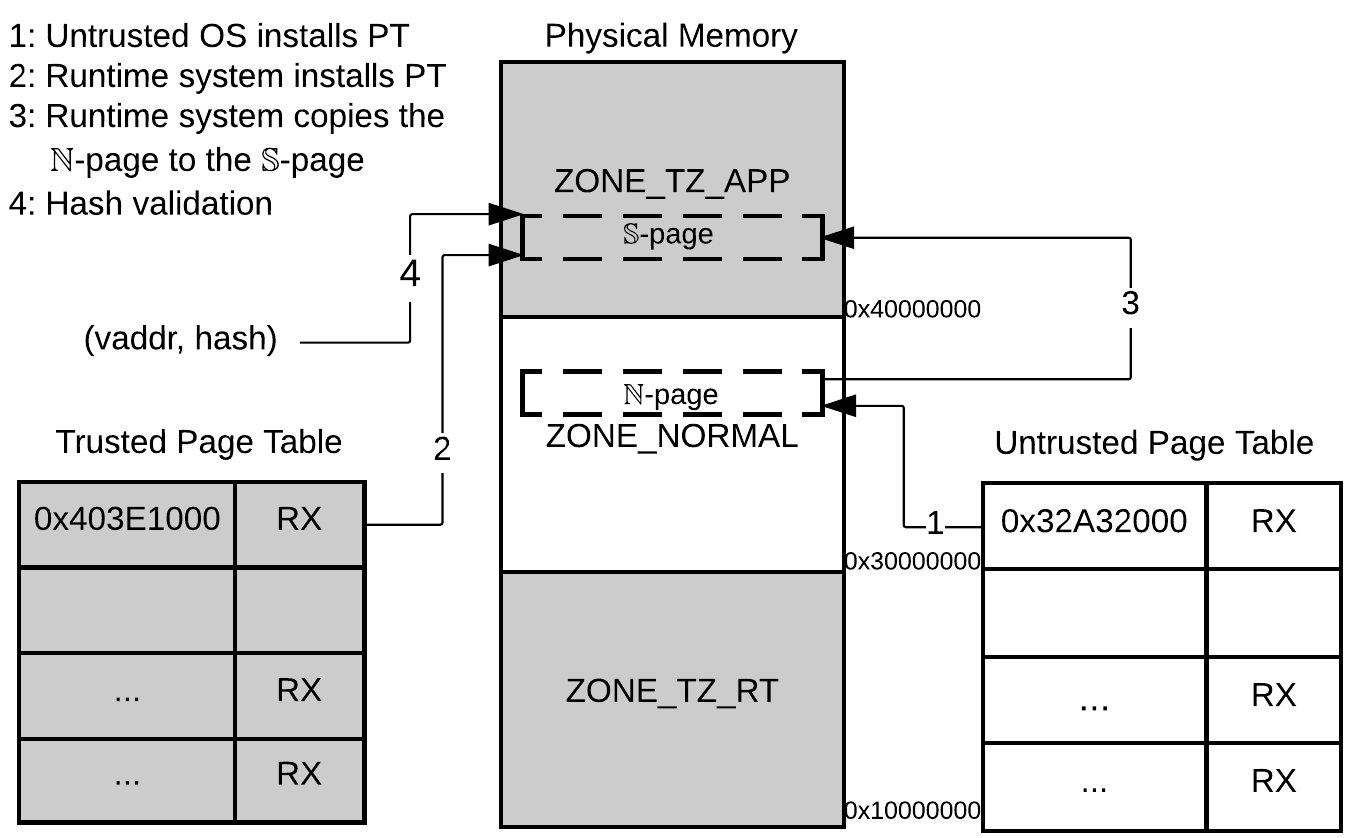}
        \caption{Page Table Update with Integrity Checking.}
        \label{fig:syncpt2}
    \end{subfigure}
    ~   
    \begin{subfigure}[t]{0.32\textwidth}
        \centering
        \includegraphics[width=1\columnwidth]{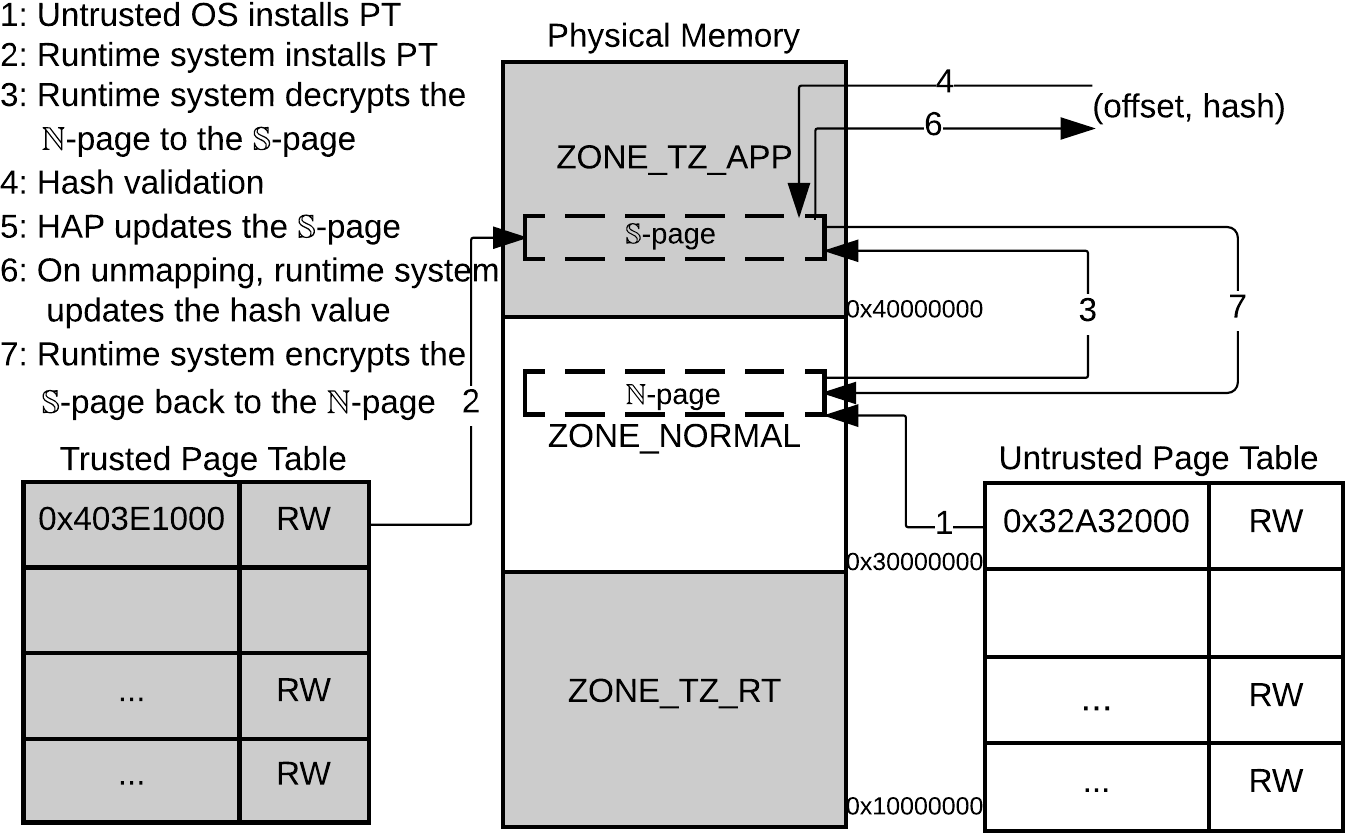}
        \caption{Page Table Update for Protected Files.}
        \label{fig:syncpt3}
    \end{subfigure}


   \caption{\label{fig:syncpt}Page Table Update.}
\end{figure*}

\subsubsection{Basic Page Table Update}
\label{sec:pagetablesyn:direct}

A page fault occurs in various situations, and \system handles page faults
differently. The most simple situation is accessing anonymous
memory, in which case the page table entries in the Linux OS and the runtime system
point to the same secure page $\mathbb{S}$. Besides duplicating the page table
entry retrieved from {\tt task\_shared}, the run time system must first make
sure that the mapped page is within the memory region defined in
\texttt{ZONE\_TZ\_APP}. We illustrate this basic page fault handling in
Figure~\ref{fig:syncpt1}.

\subsubsection{Page Table Update with Integrity Check}
\label{sec:pagetablesyn:integrity}

Different from anonymous memory, accessing memory backed by a file needs
assistance from the OS to load the file contents to memory. As the OS may be
hostile, besides ensuring that the translated pages are within the range of
\texttt{ZONE\_TZ\_APP}, the runtime system also verifies the integrity of
loaded contents. This section describes page fault handling on the memory
regions backed by executable images.

We take loading a code page as an example in Figure~\ref{fig:syncpt2}.
When a prefetch abort happens,
the Linux page fault handler will eventually call \texttt{do\_read\_fault},
which locates the physical page $\mathbb{N}$ caching the corresponding code page (Step 1).
In this context, a new secure world page $\mathbb{S}$ from \texttt{ZONE\_TZ\_APP} is allocated,
and the physical addresses of both $\mathbb{N}$ and $\mathbb{S}$ pages
are saved in \texttt{task\_shared}. 
With this shared information, 
the runtime system first ensures 
that $\mathbb{S}$ page is actually a fresh page from \texttt{ZONE\_TZ\_APP}.
Then, the runtime system 
installs a new page table entry in the trusted page table (Step 2),
copies the $\mathbb{N}$ page to the $\mathbb{S}$ page (Step 3)
and verifies the integrity of the copied page (Step 4).
Note that verification is performed on $\mathbb{S}$ pages,
therefore, \system is resilient to 
\texttt{TOCTTOU} (Time Of Check To Time Of Use) attacks.

The described page table update with integrity check is the low level primitive
for ensuring \emph{load time program integrity}. \system enforces
such checking on all the memory segments of type \texttt{PT\_LOAD} in the ELF
program images, including executable and dynamic libraries. 
In the following, we provide details on verifying the integrity of
program images.


\paragraph{Verifying Executable Integrity.}
The Runtime system maintains a list of hash values in the format of
\texttt{(vaddr, hash)}, which is initialized according to the bundled manifest
(see Section~\ref{sec:manifest}). Once a page fault occurs in the covered
range, the runtime system installs a secure page table entry as mentioned
above.
If validation is failed, the
runtime system immediately terminates the process by sending an \texttt{\_exit}
system call to the OS. We note that such validation is a one-time effort, so
it does not influence execution performance at run time when the program is
warmed up.

\paragraph{Verifying Shared Library Integrity.}
Different from executables, shared libraries are position
independent. To verify pages loaded for shared libraries, the runtime system
maintains a system wide \texttt{(offset, hash)} list for all
shared libraries frequently used. When a shared library image is mapped in the address space, the runtime system 
obtains the loaded base address
\texttt{baseAddr} by monitoring the return value of the
\texttt{mmap} system call. Then, the integrity of the loaded page is
verified at address \texttt{(baseAddr + offset)}.

\subsubsection{Page Table Update for Protected Files}
\label{sec:pagetablesyn:file}
This section describes page fault handling of data files.
Different from executable image,
the purpose of protecting data file is to prevent OS from accessing clear-text
contents. Therefore, the runtime system further employs encryption technique 
in handling this type of page fault.
\system allows developers to differentiate data files based on their the sensitivity levels.
Only sensitive files that are specified in a manifest bundled with the application (see
Section~\ref{sec:manifest} for details) are protected.

Before elaborating page fault
handling when accessing protected files, we first describe how \system manages
them at high level.  All the operations accessing these files are
transparently transformed into memory mapped I/O. To correctly map file
descriptor offsets to virtual addresses, preceding pages of a file are reserved
for meta-data. This includes the real file length, time stamp of the last
access, along with hash values of the real data pages. These preceding meta pages
are protected by a per-application AES key that is provided by the manifest.



As shown in Figure~\ref{fig:syncpt3}, when accessing a non-present page of a
protected file, the runtime system decrypts the cipher-text $\mathbb{N}$ page
loaded by the OS, and writes it into a secure $\mathbb{S}$ page (Step 3), which is also
allocated by the OS and verified by the runtime system. After that, the hash
value of the page is calculated and validated (Step 4). When unmapping this page, the
runtime system recalculates and stores the hash value of the updated
$\mathbb{S}$ page (Step 6), and then encrypts it into the original $\mathbb{N}$ page (Step 7),
which is finally written to the permanent storage by the OS. The encryption
key, again, is the per-application AES key that comes with the manifest.

\subsection{Intervening System Calls}
\label{sec:intervention}

Two problems are raised when a system call is forwarded to the OS. First, due
to isolation, the OS kernel cannot access the address space of a shadow HAP,
while some system call services rely on input data from user space. Second,
the results returned by the OS are not trusted, which may lead to potential
attacks. The runtime system coordinates the intervention between an HAP and
the OS, provides the OS with essential service request data, and verifies the
responses from an untrusted OS. For critical system services that cannot be
served by the OS (\eg~random number generator), the runtime system implements
them inside the secure world, which is discussed in
Section~\ref{sec:localservice}.

\subsubsection{Adapting System Calls}
\label{adapt_syscall}
Memory isolation changes the way that the OS manages and accesses the memory
of an HAP. Without the runtime system acting as an intermediator, it is
impossible for the OS to access application data containing system call
requests. We follow existing marshaling techniques available on x86 platform,
in which system call parameters are adapted in a world-shared buffer. This
allows the OS to have temporary access to system call parameters. Besides
this, there still remains challenges that are specific to our design. This
section briefly reviews existing marshaling technique, and then describes our
specifics.



\paragraph{Parameter Marshaling.}
Most system call parameters are scalar in that they contain values instead of
pointers to memory, for instance, \texttt{close}, \texttt{getpid}, and
\texttt{\_exit}. The runtime system forwards them directly without any
modification.

However, complex system calls allow a process to pass in a pointer so the
kernel can read data from or write result to a user space buffer. For example,
the \texttt{open} system call passes in a buffer containing the file path as a
pointer. As the OS cannot access user space data, the runtime system marshals
them into a world-shared buffer, and adjusts the parameters accordingly. The
system call service works on the marshaling buffer. After it completes, the
runtime system copies back the results into the original user buffer if
necessary.
More complex system calls, such as
\texttt{ioctl} and \texttt{fcntl}, have different behaviors according to subcommands.
A marshaling code for each request/cmd must be prepared separately according to the
specifications of the subcommands.



\paragraph{Signal.}
In signal handling, a signal delivery allows an untrusted OS to resume  user
space code at arbitrary location, thus compromising control flow integrity of
a shadow HAP. In addition, \texttt{setup\_frame} needs to manipulate the
process's stack to craft signal information and return code, while the OS has
no privilege to do so.

\system addresses these problems by both hacking the OS code and supporting in
the runtime system. Specifically, when a signal is registered, the runtime
system inserts the handler address into the \texttt{task\_private} structure
of the shadow HAP. When a signal is caught by the OS, a reserved page in the
marshaling buffer is used by \texttt{setup\_frame} to set up a separate user
mode stack specifically for signal handling. At the same time, the intended
return address for signal handler is placed in \texttt{task\_shared}. When the
runtime system resumes, it first verifies that the address has been registered
and that the \texttt{pretcode} on the signal stack is
correct\footnote{\texttt{pretcode} points to a piece of code calling the
\texttt{rt\_sigreturn} system call on sigpage. This piece of code is common to
all the processes.}. If so, the signal stack is copied to an unused virtual
address backed by a secure page\footnote{\system reserves configurable number
of secure pages specifically for this purpose.}. Then the hardware context of
the normal control flow is saved in a temporary structure in
\texttt{task\_private}, and is replaced with the signal's hardware context.
When the signal handler returns by issuing the \texttt{rt\_sigreturn} system
call, the hardware context of the normal control flow is restored.


\paragraph{Futex.}
Fast userspace mutex (futex) is another interesting kernel service that
conflicts with process isolation. In Linux, a futex is identified by a
four-bytes memory shared among processes or threads. It acts as a building
block for many higher-level locking abstractions such as semaphores, POSIX
mutexes, and barriers. If a thread fails to acquire a lock, it passes the
lock's address along with its current value to a \texttt{futex} wait
operation. This \texttt{futex} operation will block the thread if and only if
the value in lock's address still matches the value it received. The blocked
thread resumes when another thread releases the lock by issuing a
\texttt{futex} wake operation, which unblocks all the threads waiting on a
specific lock. Obviously, the \texttt{futex} system call needs to read the
value of the lock which is in the user space of an HAP.

We observe that a thread never waits for more than one futex at a
time\footnote{A blocked thread can never issue another \texttt{futex} wait
operation.}. Therefore, we hack the \texttt{futex} system call to always read
from a fixed memory location in the marshaling buffer. Each time a
\texttt{futex} wait operation is issued, the runtime system synchronizes the
current futex value to that fixed address. In \system, we further handle a
futex shared across processes by maintaining a system wide map that keeps
physical addresses of involved memories. The runtime system queries this map
to synchronize futex updates to different processes.


\subsubsection{Defeating Iago Attack}
\label{sec:iago}
As disclosed in \cite{iago}, a compromised OS could subvert an HAP by
manipulating the return values of system calls, thus leading to Iago attacks.
For example, when an HAP requests a new memory region through the \texttt{mmap}
system call, it expects that the returned region is disjoint with any existing
mapping in the process's address space. However, a compromised OS could return
an address that overlaps with the process's stack. Without proper checking on
the return values, the following write on the new region would smash the stack
and the process can be coerced into executing a return-oriented
program~\cite{rop}.

With the runtime system sitting in-between the shadow HAP and the untrusted
OS, it is straightforward to address known Iago attacks by interposing the system
call interface and checking their results.
For a known Iago attack, we need a specification for that particular system service.
Here, we take the \texttt{mmap}
system call as an example. Every return address of the \texttt{mmap} or
\texttt{brk} system system call is compared with the current memory mapping.
If an overlap is found, the HAP is immediately killed. The runtime system
collects current memory mapping in three ways. First, the range of stack is
obtained from current \texttt{sp}, because the stack spans from \texttt{sp} to
the top of user space virtual memory. Second, heap limit can be monitored by
examining return values of the \texttt{brk} system calls. Finally, the
return value of each successful \texttt{mmap/munmap} system call is recorded.


\subsection{Internal Exception Handling}
\label{sec:localservice}

In this section, we list security-critical exceptions that are handled
directly inside the runtime system. Forwarding them to the OS would leak user
data or lead to security breach.

\paragraph{Floating Point Computation.} ARM architecture supports hardware
floating point calculation by Vector Floating-Point (VFP) architecture
extension. VFP introduces a set of registers and instructions specific for
floating point calculations. The access to them is controlled by a privileged
register \texttt{FPEXC}. In Linux, when a program accesses VFP for the first
time, an \texttt{UNDEFINED} exception is raised and the kernel is responsible
for enabling VFP support for this program. To support multiple processes
accessing VFP concurrently, the kernel maintains a VFP context for each
process in its kernel stack. This design obviously leaks user data contained
in VFP registers to kernel. In \system, the runtime system duplicates the code
handling VFP from the Linux OS.
More specifically, the runtime system maintains a VFP context in the secure memory for each HAP that requires VFP calculation, and clears VFP registers whenever switching to the ordinary OS.

\paragraph{Random Number Generator.}
The Linux pseudo-Random Number Generator (LRNG) is the main source of
randomness for many cryptographic applications, such as OpenSSL. Linux
provides LRNG service by exposing \texttt{/dev/(u)random} devices to
applications. Since using weak random values is a catastrophe for 
cryptographic systems, and an untrusted OS should not know the key materials
used in the application, \system provides a trusted RNG service in the
secure world. Specifically, the runtime system maintains a list of file
descriptors that correspond to opened \texttt{/dev/(u)random} devices. Read
operations on these descriptors are intercepted such that trusted random
values are directly provided. The runtime system readily utilizes the on-board
hardware random number generator RNG4 to generate strong random numbers.



\begin{figure}[t]
        \centering
        \includegraphics[width=0.6\columnwidth]{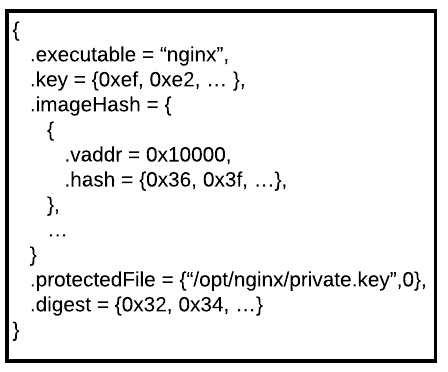}
        \caption{A sample manifest. Note that {\tt protectedFile} specifies the file names that \system needs to protect; {\tt digest} indicates the signature of this manifest.}
        \label{fig:manifest}
\end{figure}

\subsection{Manifest Design}


\label{sec:devicekey}
\label{sec:manifest}
\label{sec:keymanagement}

As mentioned in prior sections, each HAP is bundled with a manifest that
provides metadata for the security features.
We design a manifest to contain the following -- a per-application secret key,
the integrity metadata of the application (\ie~the \texttt{(vaddr, hash)}
list), and a list of file names that should be cryptographically protected.

Since the manifest is stored on a local persistent storage which can be
accessed by the OS, we design two mechanisms to ensure its security. First, we
encrypt the per-application secret key using a per-device public key.
Therefore, only the runtime system which has access to the per-device private
key is able to decrypt it. Second, to ensure the integrity of the manifest, we
append a digital signature calculated on the content of the manifest using a
per-device private key. In a real deployment, we note that per-device
public/private key pairs used for encryption and signature should be
separated. In the presentation of this paper, we refer to them as a single key
pair for simplicity. Figure~\ref{fig:manifest} shows a
C-data-structure-equivalent of a sample manifest we used to safeguard the
Nginx web server.

\begin{figure}[t]
        \centering
        \includegraphics[width=0.5\columnwidth]{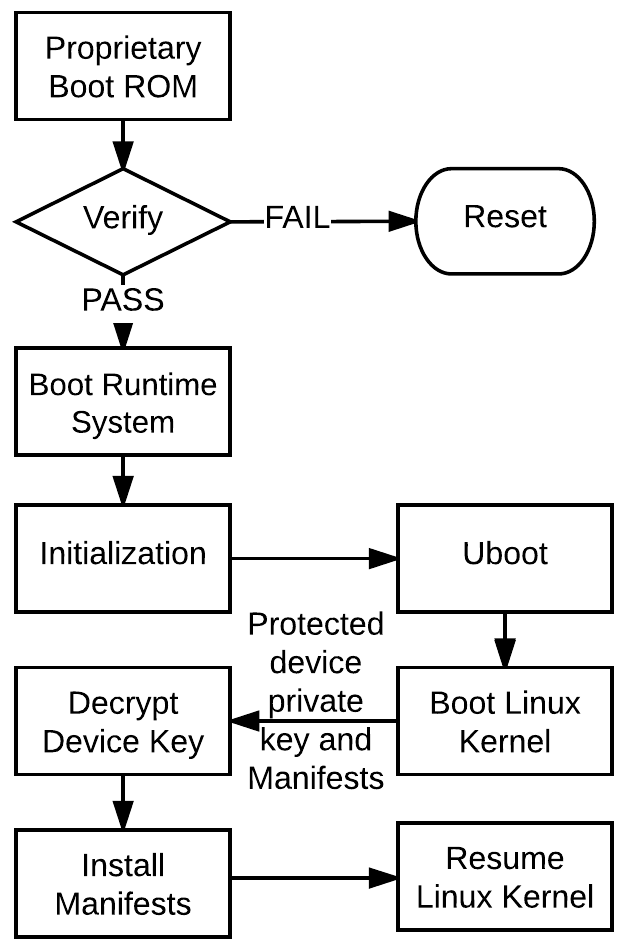}
        \caption{The boot sequence of \system. The components on the left and right indicate the operations in the secure and normal world, respectively.}
        \label{fig:boot}
\end{figure}

\section{Implementation}
\label{implementation}

We have implemented \system on a {\tt Freescale i.MX6q} ARM development board that integrates an ARM Cortex-A9 MPCore processor, 1GB DDR3 DRAM and 256KB iRAM. As is discussed in the section above, \system involves operations on both the normal and secure worlds. In this section, we therefore describe our implementation details in turn.



\subsection{Normal World}

In the normal world, we made the following changes to a Linux OS distribution with kernel version 3.18.24. (i) We added kernel parameter \texttt{tz\_mem=size@start} which indicates the memory region used for HAPs, \ie~\texttt{ZONE\_TZ\_APP}. (ii) We changed zone-based allocator to ensure that the pages designated to shadow HAPs must be from the memory region indicated by \texttt{tz\_mem=size@start}. (iii) We added a \texttt{tz} flag to \texttt{task\_struct} in order to make the OS capable of distinguishing HAPs. (iv) We implemented a new system call \texttt{tz\_execve} in order to start an HAP in Linux. (v) We changed the control flow of \texttt{ret\_to\_user} and \texttt{ret\_fast\_syscall}, so that the Linux OS can pass the execution back to a corresponding shadow HAP instead of a zombie HAP. (vi) We hooked the page fault handler so that it can prepare page table update information for the runtime system. (vii) We modified the code handling signals in order to set up a signal stack in the marshaling buffer and make it ready for an HAP. In total, these changes introduce about 300 LOC to the Linux kernel.

\subsection{Secure World}
\label{sec:securekernelimple}

In the secure world, we implemented the aforementioned runtime system with about 4.5K LOC of ANSI C and 0.8K LOC of assembly. In addition, we implemented a secure boot mechanism to guarantee the integrity of \system. Figure~\ref{fig:boot} describes how we implemented a secure boot for \system. 

Using High Assurance Boot (HAB), a proprietary boot ROM first loads the image of the runtime system. Then, it performs a verification and examines the integrity of the image. After passing the integrity check, the runtime system starts, using TZASC to configure the access policy of memory regions \texttt{ZONE\_TZ\_RT}, \texttt{ZONE\_NORMAL}, and \texttt{ZONE\_TZ\_APP}. To guarantee the policy cannot be maliciously altered, the runtime locks the configuration. 
As a result, further modifications to the policy requires system reboot.

After the success of initialization, the runtime system loads uboot binary~\cite{uboot} which further boots the Linux system implemented above. The Linux system runs in the normal world where it retrieves the manifest as well as the public/private key pair stored on the persistent storage. Note that, our implementation encrypts the public/private key pair in advance using the 256-bit Zeroizable Master Key (ZMK) stored on {\tt Freescale i.MX6q} board. This ensures the key pair is not disclosed to the Linux in plaintext. We believe this implementation is a common practice for many device manufacturers~\cite{knoxworkspace}.

To facilitate the secure boot, the Linux system passes the manifest and public/private key pair to the runtime system which further decrypts the key pair and installs the manifest. With this process completion, the runtime passes the execution back to the Linux system.

\section{Evaluation}
\label{sec:eval}

In this section, we evaluate \system by conducting extensive experiments. Using microbenchmarks, we first explore the impact of \system upon primitive OS operations. Second, we quantify the overhead of I/O operations imposed by \system. Last, we run real world applications and study the overall performance overhead introduced by our system. 
We conducted the aforementioned experiments on a {\tt Freescale i.MX6q} board running both native Linux and our \system. We treated the performance observed from native Linux as our baseline and compared it with that observed from \system.

\newcommand{\tabincell}[2]{\begin{tabular}{@{}#1@{}}#2\end{tabular}}

\begin{table}
\centering
\begin{adjustbox}{max width=\columnwidth}
\begin{tabular}{|p{2.58cm}|c|c|c|c|c|} \hline
&\multicolumn{2}{c|}{Latency ($\mu s$)} & \multicolumn{3}{c|}{Overhead} \\ \hline

\textbf{Test case}&\textbf{\tt Linux}&\tabincell{c}{\textbf{\tt Trust}\\ \textbf{\tt Shadow}}&\tabincell{c}{\textbf{\tt Trust}\\ \textbf{\tt Shadow}} & \textbf{\tt InkTag}&\tabincell{c}{\textbf{\tt Virtual}\\\textbf{\tt Ghost}}\\ \hline
{\tt null syscall} & 0.7989& 1.6048 & 2.01x & 55.80x & 3.90x\\ \hline
{\tt open/close} & 29.2168 & 40.7886 & 1.40x & 4.83x & 7.95x\\ \hline
{\tt mmap (64m)} & 559.0000 & 784.0000 & 1.40x & 4.70x & 9.94x\\ \hline
{\tt pagefault} & 4.7989 & 7.9764 & 1.66x & 1.15x & 7.50x\\ \hline
{\tt signal handler install} & 1.6257 & 3.8294 & 2.36x & 3.24x & {-}\\ \hline
{\tt signal handler delivery} & 51.6111 & 57.0349 & 1.11x & 1.61x & {-}\\ \hline
{\tt fork+exit} & 987.0000 & 2328.6000 & 2.36x & 4.40x & 5.74x\\ \hline
{\tt fork+exec} & 1060.3333 & 2509.0000 & 2.37x & 4.20x & 3.04x\\ \hline
{\tt select (200fd)} & 15.0707 & 18.8649 & 1.25x & 3.40x & {-}\\ \hline
{\tt ctxsw 2p/0k} & 30.3700 & 32.7100 & 1.08x & {-} & 1.41x\\ \hline
\end{tabular}
\end{adjustbox}
\caption{LMbench micro-benchmark results.}
\label{tab:lmbench}
\end{table}

\subsection{Microbenchmarks}

Using LMBench~\cite{lmbench}, we study the overhead imposed to basic OS operations. More specifically, we ran various system services present in Table~\ref{tab:lmbench} against both native Linux and \system. To minimize the noise involved during our experiment, we ran each benchmark with 1,000 iterations and took the average as our measures. We also compared the overhead imposed by \system with that introduced by {\tt InkTag}~\cite{hofmann2013inktag} and {\tt VirtualGhost}~\cite{CriswellDautenhahnAdve2014}. This is because both share the same goal with our system although they are designed specifically for x86 architecture and not applicable to IoT devices typically embedded with ARM processors.

Table~\ref{tab:lmbench} shows the results indicating the overhead that \system imposes to various system services. In addition, it presents the overhead introduced by {\tt InkTag} and {\tt VirtualGhost}. Note that we did not run experiments on these two systems. Rather, we obtained their overhead measures from the articles previously published in~\cite{hofmann2013inktag, CriswellDautenhahnAdve2014}. 

First, we observe that \system introduces considerable overheads to individual operations. Most notably are {\tt fork+exit}, {\tt fork+exec} and {\tt signal handler install}, all of which increase overhead by about 2.36x. The high overhead introduced by the first two services is mainly due to the fact that \system optimizes the OS to populate all the marshaling buffer in one go when creating a new thread. And, the high overhead imposed by {\tt signal handler install} results from copying a signal stack from the page in the normal world to one in the secure world.

Second, across most test cases shown in Table~\ref{tab:lmbench}, we observe that the overhead imposed by \system are relatively lower than that introduced by {\tt InkTag} and {\tt VirtualGhost}. The reason is, {\tt InkTag} and {\tt VirtualGhost} require additional CPU cycles to communicate with Virtual Machine (VM) or execute the code instrumented to kernel, whereas \system does not rely on VM nor instrument large amount of code to kernel. Compared with {\tt InkTag}, we also observe that \system imposes high overhead to system service {\tt pagefault}. This is because \system needs to perform additional page copy or zeroization, which is not required for {\tt InkTag}. 

While the overhead shown in the table appears large, it should be noted that, this does not imply that \system jeopardizes the performance of applications under protection. In fact, applications are significantly less sensitive to system services. As we will show later in the section, \system imposes only negligible performance overhead to application execution.

\begin{figure}
        \centering
        \includegraphics[width=1\columnwidth]{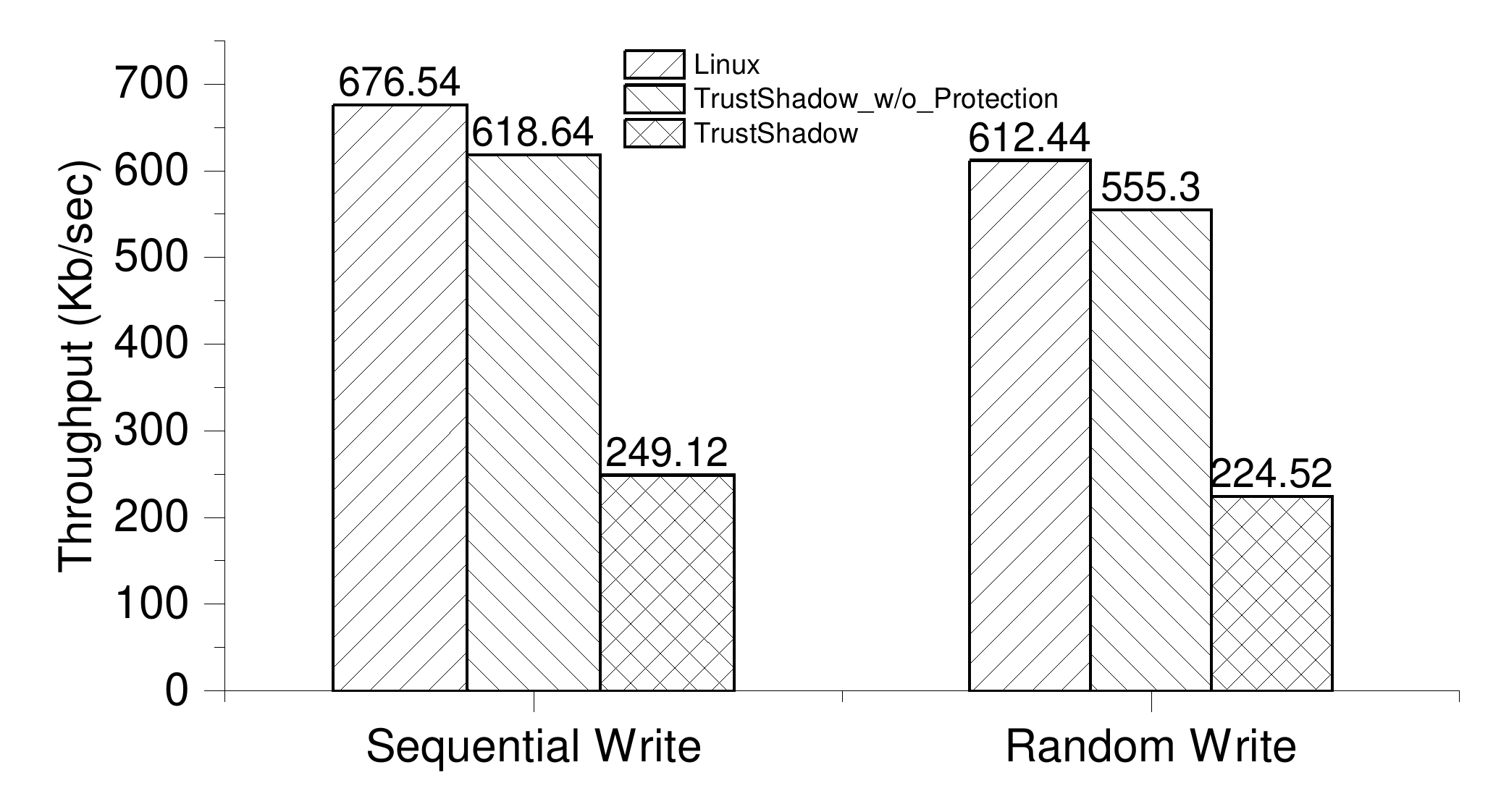}
        \caption{File I/O performance as measured by sequential and random write.}
        \label{fig:file}
\end{figure}
\subsection{File Operations}

To quantify the overhead imposed to I/O throughput, we conducted an experiment by 
using Sysbench~\cite{sysbench} in different modes. As is discussed earlier,
\system allows developers to designate whether or not to protect a particular
file. Thus, we did this experiment with and without file protection enabled.

We prepared 128 files, each of which has 8Mb, and tested both sequential
write and random write. To minimize cache effects and best reflect the actual
I/O performance, we conducted experiments with write-through mode, and forced
a call to \texttt{fsync()} after each write operation.

Figure~\ref{fig:file} shows the results. Because caching is disabled,
sequential write did not exhibit significant advantage over random write. In
both cases, the slowdown introduced by \system without file protection is
about 1.09x, while the slowdown by \system with file protection is about
2.71x. This is due to the fact that \system with file protection enabled
involves heavy encryption and hashing computations when it synchronizes pages 
to persistent storage. Note that the difference between \system with and without file
protection solely results from cryptographic operations. As a result,
employing a more efficient cryptographic engine is a straightforward way
of improving file I/O performance.

\begin{figure}
        \centering
        \includegraphics[width=1\columnwidth]{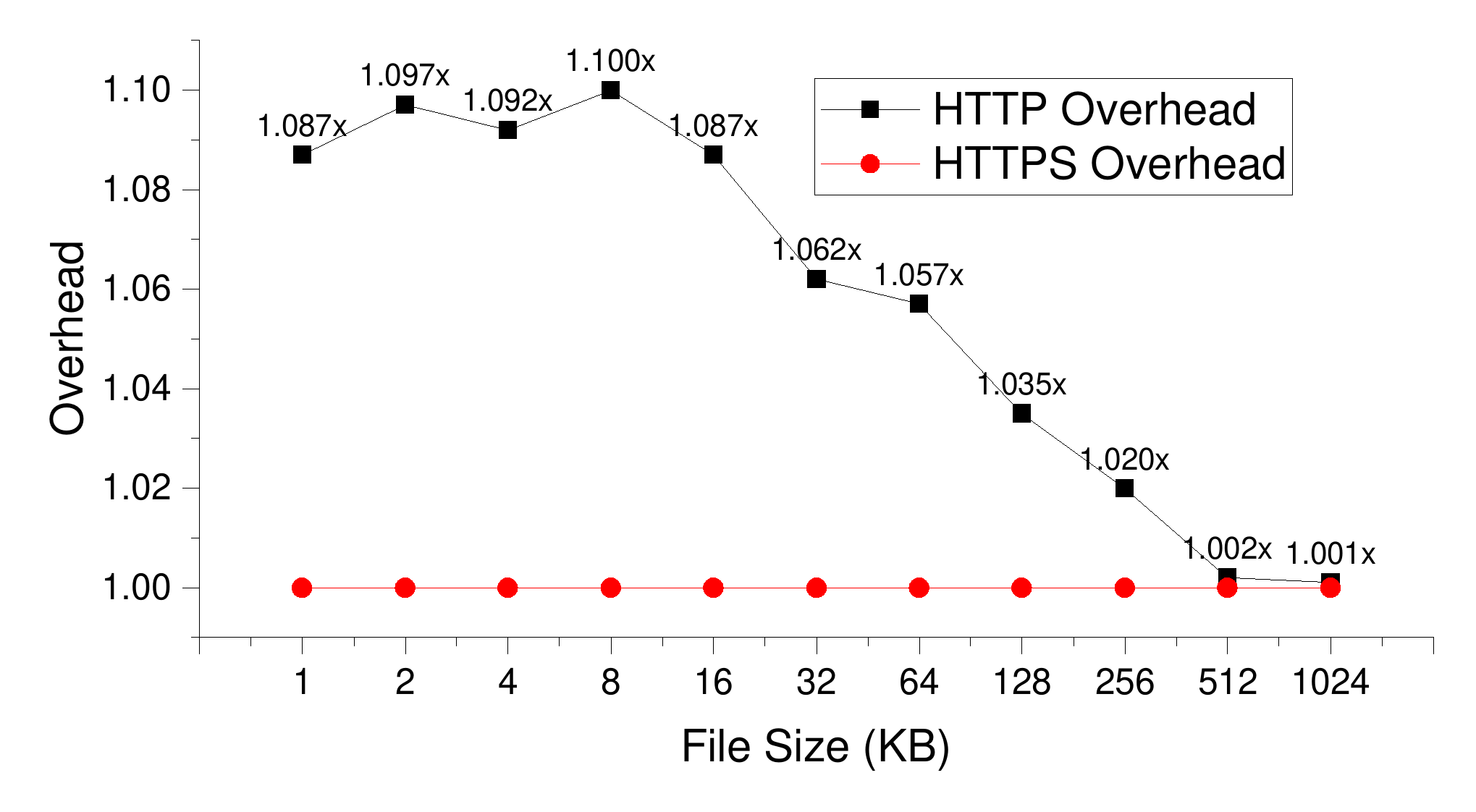}
        \caption{Throughput overhead imposed by \system across HTML responses in different sizes.}
        \label{fig:nginx}
\end{figure}

\subsection{Embedded Web Server}







To study the impact of \system upon real-world applications, we mimic an embedded web server running on an IoT device. More specifically, we ran the Nginx web server with version 1.9.15 on our testbed against both native Linux and \system. We configured Nginx to respond with HTML file in different sizes through both HTTP and HTTPs. To quantify the server throughput, we utilized the Apache benchmark~\cite{ab} on another machine to connect to the server. We configured the benchmark program to create 10 connections simultaneously sending 10,000 HTTP requests. This experiment setting allows us to overwhelm the server and thus compare the throughput variation from the viewpoint of an end user.


Figure~\ref{fig:nginx} shows the throughput overhead of Nginx. Regarding to the requests through HTTP, we observe that, \system downgrades the server throughput by about 6\% $\sim$ 10\% when a client requests a file in a relatively small size. However, this throughput downgrade is alleviated when the requested file increases. As is shown in Figure~\ref{fig:nginx}, the server throughput drops only by 2\% when a client requests a file with a size more than 256 KB. 
Regarding latency, we measured the 95\% percentile in each experiment, and found almost no latency
overhead. The raw HTTP performance measurements can be found in Appendix~\ref{app:http}.
In addition, we discover \system introduces nearly no overhead regardless of whether the size of the file requested varies when HTTP traffic is carried through TLS. The reason behind these observations is that the overhead imposed by \system is overwhelmed by intensive I/O and computationally intensive cryptographic operations, respectively.



\subsection{Data Analytics}

We also evaluate the performance overhead when performing data analysis using machine learning techniques. This kind of applications is popular for many edge computing platforms.
To evaluate the impact of \system upon such applications, we chose two data analytic applications to emulate face recognition and image classification. To be more specific, we ran an DLIB image classification program~\cite{dlib} powered by deep neural networks against 60,000 images from the MNIST database~\cite{MNIST}, as well as a face recognition program against 10 images with human faces from the DLIB toolkits.

In the first experiment, we observed that \system classifies 60,000 images in 998 seconds, which is only 4 seconds higher than performing the same classification on native Linux. In the second experiment, we observed that \system recognizes faces in 163.271 seconds, whereas native Linux finishes the same task in 162.591 seconds. The observations from both experiments again indicate \system introduces only negligible overhead to IoT devices.

\section{Discussion and Future Work}
\label{sec:discussion}

In this section, we first summarize how \system defeats OS level attacks to an
HAP. Then, we analyze the security of \system, quantify its TCB, and discuss
the remaining attack surface. Finally, we discuss future work.



\paragraph{HAP Security.}
\system protects an application from three aspects.
(1) With a mechanism to verify the integrity of a program image, 
an attacker cannot manipulate application code/data at \emph{load
time}. (2) With the isolation of resources and introspection mechanism, an attacker cannot interfere with HAP execution at \emph{run time}. The only user memory that
the OS can access is the marshaling buffer.
(3) With a cryptographic mechanism encrypting files and signing meta data, attackers can no longer read a file under protection or make any modification to it.
\paragraph{Runtime System Security.}
The protection above is established on the basis of the correctness and
robustness of the runtime system. Our design guarantees the security of the runtime system from three aspects.
First, our design ensures the integrity of the runtime system at load time because 
the hash of the verification public key is burned in the chip's fuses and 
HAB uses this key to verify the signature of the runtime system image
before loading it. 
Second, our design reduces the attack surface of the runtime system because the runtime system itself is loaded into a secure
physical region which the ordinary OS cannot manipulate. Third, our design raises the bar for exploiting the runtime system. This is due to three reasons. (1) An application
must undergo critical security reviews before being authorized to run as an
HAP (\ie~providing it with a manifest with manufacture signature). (2) Even if an HAP has vulnerabilities that may be exploited to execute arbitrary code, it only runs with user privilege. (3)
The interface exposed by the runtime system is narrow, because it simply forwards most exceptions to the OS. Also,
the small code base of runtime system makes it possible for formal verification.

\paragraph{TCB Size.}
To demonstrate the security of \system quantitatively, we identify the TCB of our system, and compare its size with x86 alternatives.

Ultimately, all the code in the user's TCB must be trusted. Therefore, like all the other works in this line, user application is included in the TCB. The size of user application is highly dependent on its functionality and complexity. We rely on code review to achieve trust for this part of TCB.

The runtime system maintains the execution environment for an HAP, and thus must be included in the TCB.
As mentioned earlier, our runtime system has only about 5.3K LOC, which we believe is small enough for manual review or formal verification.
In comparison, previous x86 works have their own privileged code that must be trusted.
Hypervisor-based solutions~\cite{chen2008overshadow,ChenZhangChenEtAl2007,YangShin2008,hofmann2013inktag} include the whole hypervisor in its TCB, bloating their TCBs by several hundreds of thousands of lines of code. Although thinner hypervisors exist~\cite{trustvisor}, we are not aware of any similar system built on top of them.
Haven~\cite{BaumannPeinadoHunt2014} includes LibOS, a large subset of Windows in its TCB, resulting a TCB of millions LOC.
VirtualGhost~\cite{CriswellDautenhahnAdve2014} includes about 5.3K LOC for their run-time system and LLVM passes. This is the only solution that has comparable TCB with \system.




\losseparagraph{Remaining Attack Surface.}
To minimizes TCB, the runtime system does not implement system services
itself, but relies on the OS. With full control of process scheduling, the OS
can easily launch DoS attacks to an HAP. Similarly, to
start an HAP, the OS may choose to invoke the normal \texttt{execve} system
call instead of \texttt{tz\_execve}. However, the process is executed in the
normal world, so it cannot access cryptographically protected files.

Another
concern is the manipulation of manifest files. If a manifest file can be
forged, integrity checking of the corresponding executable image is bypassed.
As a result, arbitrary code can be loaded in the secure world. We address this
problem by signing the manifests using a per-device private key.
When a vulnerable program is updated, the corresponding manifest should be updated
as well. A roll-back attack happens when an attacker executes the vulnerable version
of the program with an older manifest.
To prevent this from happening, one of our future work is to add a version number field
in the manifest, and periodically communicate a list showing the updated version numbers of trusted programs between the runtime system and a remote server.



Last but not least, side channel attacks have been developed to extract
information across processes in an OS, or even virtual
machines~\cite{YaromFalkner2014,ZhangJuelsReiterEtAl2014,cachestroage,armageddon}.
For example, in~\cite{cachestroage}, the authors introduced a cache storage
channel that exploits the inconsistence of cache and physical memory data, and
infers victim's behaviors in TrustZone by checking if the constructed
inconsistence has been destroyed. \system's current design may be subject to
this line of side channel attacks. However, we can adopt existing techniques
to mitigate such attacks. For example, contemporary cryptographic libraries
such as OpenSSL have already been designed to resist some side channel
attacks~\cite{eliminatingtiming}.


\label{sec:securityanalysis}





\losseparagraph{Future Work.}
Many IoT devices, such as cyber-physical systems, are sometimes deployed in
an unmonitored environment. As a result, secret data stored in the DRAM chip
is subject to inexpensive physical attacks, such as cold-boot
attack~\cite{cold-boot,frost}, bus
monitoring~\cite{gogniat2008reconfigurable}, and DMA attacks~\cite{DMA2,DMA4}.
In~\cite{smartphonescoldboot,ZhangSunLouEtAl2016}, the authors proposed to
process confidential data within SoC components such as cache and iRAM,
because it is considered much harder to compromise a SoC component. In
\system, we can simply configure the region \texttt{ZONE\_TZ\_APP} to be within the
range of iRAM to make HAPs immune to physical attacks.

We have tested this idea on our experiment board, which integrates a 256KB
iRAM. We successfully ran a small program that generates a 2048-bit RSA
key-pair with OpenSSL. As the capacity of iRAM is usually limited, this
enhanced security feature could only be activated for low footprint programs.
One of our future work is to ``extend'' iRAM by introducing another level of
virtual memory -- utilizing DRAM as a backup storage for encrypted iRAM pages.

Although hypervisor-based solutions are not applicable to shield trusted
applications in ARM platform as discussed in Section~\ref{related}, the idea of
cloaking memory with a privileged layer has long been studied. With a
hypervisor, trusted applications can run in the normal world, which could
greatly reduce the risk in the presence of a vulnerable runtime system. While
TrustZone is not designed to provide features such as shadow page table that a
standard hypervisor could provide, some existing work, such as
TZ-RKP~\cite{hypervision} and Sprobes~\cite{sprobes}, has the potential to
emulate these features in the secure world. Integrating the technique of
across worlds hypervision with existing shielding mechanisms remains another
topic for our future work.





\section{Conclusion}
\label{sec:conclusion}

In this paper, we have presented \system that utilizes a carefully designed runtime system to shield applications running on multi-programming IoT devices. With \system, security-critical applications on these devices can be comprehensively protected even in the face of total OS compromise. Unlike techniques previously proposed, the design of \system does not require modification to applications. As a result, security can be guaranteed without the requirement of re-engineering applications. Since \system imposes only negligible -- and occasionally moderate -- overhead to IoT devices, the protection of an application can be achieved in a lightweight manner. With an increasing number of IoT devices developed, we expect the design of \system could inspire more research in the area of IoT computing.





\section{Acknowledgments}
We would like to thank the anonymous reviewers for their insightful feedback
and our shepherd, Ardalan Amiri Sani, for his valuable comments on revision of
this paper. This work was supported by U.S. Army Research Office award
W911NF-13-1-0421 (MURI), NSF under Grant No. CNS-1422594, CNS-1505664,
CNS-1634441, CNS-1422355, CNS-1408880, SBE-1422215, and the Penn State
Institute for CyberScience (ICS) Seed Funding Initiative grant.
Any opinions,
findings, and conclusions or recommendations expressed in this
material are those of the authors and do not necessarily reflect the
views of the National Science Foundation and U.S. Army Research Office.


%
\bibliographystyle{abbrv}
\bibliography{tz,myref,tsx}  
%
%
\appendix

\section{Code Snippet of Exception Forwarding}
\label{appen:forwarding}

\begin{lstlisting}[language={armAssembler},label={lst:forwarding}, caption={Code snippet that forwards
SVC exception to the normal OS.}]
forward_syscall:
@ we start at monitor mode
@ spsr is set to SVC, which is the target mode
mov     r0, #(MODE_SVC | I_BIT | F_BIT)
msr     spsr_cxsf, r0

@ load r0 with the address of user space CPU_context
ldr     r0, =uregs
ldr     r0, [r0]

@ switch to SVC mode to set target mode's registers
cpsid   if, #MODE_SVC
ldr     r1, [r0, #CPSR]
msr     spsr_cxsf, r1

@ restore critical registers required by normal OS in SVC, ATB and IRQ modes
ldr     r1, =svc_sp
ldr     sp, [r1]
cpsid   if, #MODE_ABT
ldr     r1, =abt_sp
ldr     sp, [r1]
cpsid   if, #MODE_IRQ
ldr     r1, =irq_sp
ldr     sp, [r1]

@ back to monitor mode
cpsid     if, #MODE_MON

@ restore syscall arguments. r0-r6 for arguments, r7 for syscall number
...
@ before switching to normal OS, change securiry state to non-secure
pop     {r4}
eor     r4, r4, #NS_BIT @ Toggle NS bit
mcr     p15, 0, r4, c1, c1, 0
isb

@ branch to 0xFFFF0008, which is the offset for the SVC exception handler
mov     r8, #0x8
movt    r8, #0xFFFF
movs    pc, r8
\end{lstlisting}

\section{Raw HTTP Performance Measurements}
\label{app:http}

\begin{table}[h]
\centering
\begin{adjustbox}{max width=\columnwidth}
\begin{tabular}{|c|S[table-format=3.2, table-figures-uncertainty=1]|S[table-format=3.2, table-figures-uncertainty=0]|c|c|} \hline
\multirow{2}{*}{\textbf{File size}} & \multicolumn{2}{c|}{Throughput (Requests/second)}  & \multicolumn{2}{c|}{95\% Percentile (ms)} \\  \cline{2-5}
& {Linux} & {TrustShadow} & {Linux} & {TrustShadow} \\ \hline
1KB & 652.50  & 600.46 & 17 & 17 \\ \hline
2KB & 622.83 & 567.96 & 18 & 18 \\ \hline
4KB & 621.72 & 569.37 & 16 & 18 \\ \hline
8KB & 603.75 &  548.74 & 17 & 19 \\ \hline
16KB & 531.15 & 488.64 & 19 & 21 \\ \hline
32KB & 433.68 & 408.25 & 24 & 25 \\ \hline
64KB & 316.64 & 299.57 & 33 & 36 \\ \hline
128KB & 220.73  & 213.25 & 48 & 49 \\ \hline
256KB & 133.26 & 130.58 & 84 & 85 \\ \hline
512KB & 75.85 & 75.69 & 158 & 158 \\ \hline
1024KB & 42.68 & 42.64 & 307 & 309 \\ \hline

\end{tabular}
\end{adjustbox}
\label{tab:http}
\end{table}
\end{document}